\documentclass[aps,prb,floatfix,twocolumn,superscriptaddress,showpacs,epsf]{revtex4-1}

\usepackage[USenglish]{babel}
\usepackage{amsmath,amssymb,amsfonts}
\usepackage{epstopdf}
\usepackage{epsfig} 
\usepackage{graphicx}
\usepackage{subfigure}
\usepackage{import}
\usepackage{color}
\usepackage{url}

\allowdisplaybreaks

\usepackage[colorlinks=true,linkcolor=blue,citecolor=red]{hyperref}

\newcommand{\ca}{c^{\phantom{\dagger}}}
\newcommand{\cc}{c^\dagger}

\newcommand{\be}{\begin{equation}}
\newcommand{\ee}{\end{equation}}
\newcommand{\bea}{\begin{eqnarray}}
\newcommand{\eea}{\end{eqnarray}}
\newcommand{\ba}{\begin{eqnarray*}}
\newcommand{\ea}{\end{eqnarray*}}

\newcommand{\up}{\uparrow}
\newcommand{\down}{\downarrow}

\newcommand{\ket}[1]{|{#1}\rangle}

\newcommand{\braket}[3]{\langle{#1}| {#2} |{#3} \rangle}

\newcommand{\quave}[1]{\langle {#1} \rangle}

\newcommand{\ucd}{U_{c,dyn}}

\def\eg{\mbox{\it e.g.\ }}

\begin{document}

\title{From sudden quench to adiabatic dynamics in the attractive Hubbard model}
\author{Giacomo Mazza}
\email{giacomo.mazza@polytechnique.edu}
\affiliation{Centre de Physique Th\'eorique, \'Ecole Polytechnique, CNRS, Universit\'e Paris-Saclay, 91128 Palaiseau, France}
\affiliation{Coll\`ege de France, 11 place Marcelin Berthelot, 75005 Paris, France}

\pacs{}

\begin{abstract}
  We study the crossover between the sudden quench limit and the adiabatic
  dynamics of superconducting states in the attractive Hubbard model.
  We focus on the dynamics induced by the change of the attractive 
  interaction during a finite ramp time. The ramp time is varied in order  to track the
  evolution of the dynamical phase diagram from the sudden quench to  
  the equilibrium limit. Two different dynamical regimes 
  are realized for quenches towards weak and strong coupling interactions.
  At weak coupling the dynamics depends only on the energy injected into the system, 
  whereas a dynamics retaining memory of the initial state
  takes place at strong coupling.
  We show that this is related to a sharp transition between a weak and 
  a strong coupling quench dynamical regime, which defines the boundaries beyond
  which a dynamics independent from the initial state is recovered.
  Comparing the dynamics
  in the superconducting and non-superconducting 
  phases we argue that this is due to the lack of an adiabatic connection
  to the equilibrium ground state for non-equilibrium superconducting
  states in the   strong coupling quench regime.
\end{abstract}

\maketitle

\section{Introduction}

The sudden change of a parameter in the Hamiltonian (quantum quench) 
is the most common protocol to probe the quantum dynamics in isolated
systems.~\cite{silva_rmp} 
This has stimulated the investigation of fundamental questions in 
quantum statistical mechanics, as \eg the conditions under which a 
closed systems in an initial excited state is able or not 
to reach a thermal state.~\cite{srednicki_ETH,Rigol_ETH,greiner_thermalization_entanglement}
In various systems this has led to the observation of several
intriguing dynamical behaviors,   like long-time trapping in pre-thermal 
states~\cite{berges_prethermalization,schmiedmayer_pretherm_science,schmiedmayer_pretherm_nature,moekel_quench} 
or transitions between different dynamical regimes separated by dynamical 
critical points.~\cite{eckstein_quench,schiro2010,sciolla_biroli_jstat,sciolla_biroli_boseH,gambassi_calabrese}

The effects of a quantum quench can be
ultimately related to a  finite energy injection into the system,
which simply results from the fact that  the initial state $\ket{\Psi_i}$ is not
the ground state of the Hamiltonian ${\cal H}_f$ setting the unitary time evolution.
The opposite limit corresponds to an adiabatic process for which the energy 
injected into the system vanishes during the Hamiltonian variation.

In principle, an adiabatic process can be realized for an infinitely
slow variation of some control parameter, leading to the continuous evolution
of the equilibrium properties of the system from the initial to the final Hamiltonian.
In practice, exceptions to this expectation are known.
This is the case, for example, of gapless low-dimensional 
systems, where the energy density may increase with the system size 
preventing the adiabatic limit to be reached in the 
thermodynamic limit.~\cite{polkovnikov_adiabatic} 
Non-adiabatic dynamics may appear also for gapped low-dimensional  
systems in an ordered phase, where the reaching of the adiabatic limit 
is hampered by the vanishing critical temperature that leads to the melting 
of the ordered phase for every  small, but finite, density of
excitations.~\cite{anna_ising}
Recently, the formation of non-thermal states resulting from an 
adiabatic dynamics has
been proposed in the case in which the initial state of the system 
is not described by a pure state.~\cite{kennes_adiabatic}

In the cases in which an adiabatic dynamics sets in, the 
time evolution following a finite rate variation of the Hamiltonian 
can be conveniently used to address the crossover between the quench and  
the equilibrium phase diagrams. 
In the presence of dynamical transitions, this has been used to 
highlight the connection between dynamical and equilibrium 
phase transition. In particular, the adiabatic mapping between 
dynamical and equilibrium critical points has been demonstrated 
in mean-field like models, as the $O(N)$ model in the large $N$ limit~\cite{anna_On}  and for the Mott
transition in the repulsive Hubbard model in infinite dimension.~\cite{sandri_ramps,eckstein_ramps}
This can be understood imaging an adiabatic dynamics which
continuously explores the instantaneous evolution of  the system
ground state, so that a dynamical critical point surviving for
a slow variation of the Hamiltonian is expected to merge with the
quantum critical one in the adiabatic limit.
% thus mapping the dynamical critical point
% into the quantum critical point for sufficiently slow variation of the
% Hamiltonian.
It should be noticed that this mapping has not been so far addressed
in models beyond mean-field. In addition, even in the cases in which
the mapping holds, a departure from perfect adiabaticity is still
expected for any finite speed crossing of a quantum critical
point.~\cite{zurek_orig,kibble_orig,zurek_ising,dziarmaga_ising,polkovnikov_kz,chandran_kz}  

% Such adiabatic connection between a dynamical transition and a quantum
% phase transition breaks down in the cases in which the equilibrium
% phase diagram does not display quantum critical points and a thermal
% phase transition is realized in all the parameter space.
% A different example is given by systems not showing any equilibrium
% critical point and for which a thermal transition is realized in all
% the parameter space.

In this paper we investigate the evolution from the quench to the adiabatic limit 
in the case of the attractive Hubbard model. The model gives the simplest description 
of the crossover between a weakly and a strongly coupled superconductor.~\cite{bcs_bec_toschi}
Its dynamical properties are directly related to the investigation of
the non-equilibrium properties of highly excited superconducting
states which have recently attracted a considerable attention in relation to
experiments in  both solid state and cold  atoms
systems.~\cite{fausti_science,esslinger_afm,randall_afm,greiner_afm} 

%In contrast to the ground state properties of the final Hamiltonian
Highly excited states induced by a sudden quench of the attractive 
interaction are characterized by the melting of the superconducting order at 
both weak and strong coupling.~\cite{barankov_bcs,yuzbashyan_bcs,peronaci_dwave,tsuji_afm,werner_afm,balzer_prx,matteo_afm}
On the contrary, the superconducting properties of the ground state of
the final Hamiltonian are expected to be recovered as the injected 
energy is reduced and the adiabatic limit is approached. 

Here we use the time-dependent Gutzwiller approximation~\cite{michele_review,seibold_tdg} 
to study 
how the ground state properties are recovered  after a slow variation
(finite ramp) of the attractive interaction. 
We show the existence of a weak and a strong coupling quench regimes
for which two sharply different approaches to the adiabatic limit are observed.
At weak coupling the recovery of the superconducting order depends only on
the energy injected into the system.
On the contrary, for quenches in the strong coupling limit the dynamics 
retains memory of the initial state and the recovery of the 
superconductivity does not depend only on the energy injected into the system.

The two dynamical regimes are closely related to a weak-to-strong
coupling dynamical transition associated to different
excitations created during  the quench. 
By tracking the finite ramp evolution of the dynamical phase diagram
and comparing to the dynamics in the non-superconducting phase we
argue that the memory dependent  dynamics in the strong coupling quench regime
is due to the lack of an adiabatic  connection between the
non-equilibrium states at strong coupling 
and any equilibrium state at a finite value of the  interaction.

In the following we will introduce the model 
and the method in Sec.~\ref{sec:model}. 
The quench phase diagram obtained within the 
time-dependent Gutzwiller approach is discussed in
Sec.~\ref{sec:quench}.
Finally, Sec.~\ref{sec:ramp} reports the evolution of the 
quench phase diagram and the recovery of the adiabatic limit 
following a finite ramp dynamics.

\section{Model and Method}
\label{sec:model}
We study the attractive Hubbard model subject to a time-dependent
interaction which is varied over a time  $\tau$
\begin{equation}
  \mathcal{H}(t) = -\sum_{ij,\sigma} t^{\phantom{\dagger}}_{ij} 
  \cc_{i \sigma} \ca_{j \sigma} - U_{\tau}(t) \sum_{i} n_{i \up}
  n_{i \down}.
  \label{eq:H_attractiveU}
\end{equation}
The attractive interaction is defined as
\begin{equation}
U_{\tau}(t) = U_{i} + r_{\tau}(t) (U_f-U_i)
\label{eq:ramp}
\end{equation}
where $U_i>0$ and $U_f>0$ are, respectively, the initial and final interaction   
and  $r_{\tau}$ is a smooth ramping function interpolating between 0
and 1 over the time $\tau$. Its definition reads $r_{\tau}(t) = 1/2-3/4\cos \pi
t/\tau + 1/4 \cos^3 \pi t/\tau $  for $t<\tau$ and 
$r_{\tau}(t)=1$ for $t\geq \tau$.
The chemical potential is chosen such that the system is half-filled
with an average occupation of one electron per site $\langle n \rangle
= 1$.

The dynamics is studied using the time dependent 
Gutzwiller (TDG) variational ansatz~\cite{michele_review}
\begin{equation}
  \ket{\Psi(t)} \equiv \prod_i {\cal P}_i(t) \ket{\Psi_0(t)},
  \label{eq:ansatz}
\end{equation}
where $\ket{\Psi_0(t)}$ is a single particle wavefunction, describing
the dynamics of the delocalized quasiparticles and ${\cal P}_i(t)$ are
local projectors describing the dynamics of the local 
many body multiplets.
The dynamics of  $\ket{\Psi(t)}$ is obtained by means of the 
time dependent variational principle
$\delta \int \braket{\Psi(t)}{i \partial_t -{\cal H}(t)}{\Psi(t)} =0$.
The time-variational principle can be exactly imposed in the limit of a lattice 
with infinite coordination number $z$. In this paper we focus on this
limit  by considering a Bethe lattice with a semicircular density of states.

The imposition of the time dependent variational leads to
a set of equations of motions which fully describes the dynamics of
the system under a generic time-dependent perturbation.
In few words, the dynamics is described by a set of
coupled equations  of motion for the delocalized quasiparticles, described by
$\ket{\Psi_0(t)}$, and for the localized degrees of freedom, described
by the projectors $\cal P$. 
The projectors $\cal P$ contain all the local variational parameter 
and are practically expressed in terms of matrices $\hat{\Phi}$ of
the size of the local Hilbert space.~\cite{michele_review}
Upon imposing the time-dependent variational principle the following equations of motions are obtained  
\begin{eqnarray}
  && i \partial_t \ket{\Psi_0(t)} = \mathcal{H}_{qp} [\hat{\Phi}] \ket{\Psi_0(t)}
     \label{eq:iHstar} \\
  && i \partial_t \hat{\Phi}(t) = \mathcal{H}_{loc}(t) \hat{\Phi}(t) +
  \frac{\delta E_{kin}(t)}{\delta \hat{\Phi}^{\dagger}(t)}  \label{eq:iHloc}
\end{eqnarray}
$\mathcal{H}_{qp}$ is a BCS-like Hamiltonian which depends on the 
local variational parameters $\hat{\Phi}$ and ${\cal H}_{loc}(t)$ represents 
the local interaction terms in the Hamiltonian Eq.~\ref{eq:H_attractiveU}.
$  E_{kin}(t) = \braket{\Psi_0(t)}{\mathcal{H}_{qp}
  [\hat{\Phi}]}{\Psi_0(t)}$ 
and $ E_{pot} (t) = \text{Tr} \left( \hat{\Phi}^{\dagger}(t) {\cal H}_{loc} \hat{\Phi}(t)  \right) $ 
give, respectively, the kinetic and potential energies. The total
energy $E(t) = E_{kin}(t) + E_{pot} (t)$ is 
conserved by the unitary dynamics (Eqs.~\ref{eq:iHloc}-\ref{eq:iHstar})
if the Hamiltonian does not depend on time.
We refer to the literature for the detailed description of the
method~\cite{michele_review} and its extension to the superconducting
case.~\cite{giacomo_c60} 

As can be appreciated from Eqs.~\ref{eq:iHstar}-\ref{eq:iHloc}, within the TDG 
 the electron dynamics is described by the an effective 
single-particle problem (Eq.~\ref{eq:iHstar}) which is self-consistently 
coupled to the dynamics of the local degrees of freedom (Eq.~\ref{eq:iHloc}).
This provides a mutual feedback between the delocalized and localized 
degrees of freedom which greatly improves the description of the  non-equilibrium 
dynamics in correlated systems with respect to standard mean-field techniques.
On the other hand, similarly to the standard mean-field methods, 
the effective single particle description of the many-body problem
has an intrinsic limitation due to the lack of a finite life time of
the quasi-particles. This prevents for example 
a correct description of the long time relaxation to true thermal
states. 

Despite such a limitation, the method provides a meaningful description of the
dynamics in the prethermal regimes, with a good qualitative agreement 
with exact approaches in infinite dimensions such as the non-equilibrium 
Dynamical Mean Field Theory (DMFT).~\cite{neqDMFT_review} 
This can be appreciated by a direct comparison of the available
results in the literature.

%Despite some intrinsic limitations of the method related to an
%effective single particle description of the many-body problem
%which prevent a correct description of the long time relaxation to thermal 
%states, it has been shown that the TDG ansatz can provide a meaningful description
%of the transient dynamics in correlated systems, with a good
%qualitative agreement with the description provided by  
%exact approaches in infinite dimension 
%as the non-equilibrium Dynamical Mean Field Theory.~\cite{neqDMFT_review} 
The simplest example is given by the quench dynamics in the paramagnetic 
repulsive Hubbard model~\cite{eckstein_quench,schiro2010}.
In this case the variational ansatz Eq.~\ref{eq:ansatz} describes a 
dynamical transition as a function of the final interaction~\cite{schiro2010} 
which correctly  reproduces the trapping into quasi-stationary pre-thermal 
states observed by DMFT at both weak and strong coupling~\cite{eckstein_quench} 
before thermalization eventually occurs on a longer time scale. 

Another case is represented by the quench dynamics in the presence of long range ordered 
states in the antiferromagnetic (AFM) repulsive Hubbard model.~\cite{tsuji_afm,werner_afm,balzer_prx,matteo_afm} 
This model is related to the Hamiltonian  Eq.~(\ref{eq:H_attractiveU}) by the particle-hole
transformation $\cc_{i\up} = (-1)^{i} \ca_{i \up}$, so that in the 
particle-hole symmetric case the two dynamics are equivalent.
In this case both DMFT~\cite{tsuji_afm,werner_afm,balzer_prx} 
and TDG~\cite{matteo_afm} predict the trapping into non-thermal states 
for which a finite order parameter is observed 
despite the energy injected by the quench would predict a thermal 
melting of the initial long range order. 

For later convenience we recall in the next section the results of the quench
dynamics in the language of the superconducting order parameter for
the present attractive Hubbard model.
In the rest of the paper we consider a Bethe lattice in the 
limit $z \to \infty$ described by a semi-circular 
density of states of width $W$, $\rho(\epsilon) =
\theta(|W|-\epsilon) 4/\pi \sqrt{W^2-4\epsilon^2}$.

In the case of finite coordination lattices the use of the Gutzwiller approximation
corresponds to the neglect of the spatial correlations. 
In the presence of a long-range order we expect that the spatial correlations play
a minor role. Therefore, the following results can be expected to hold qualitatively 
for any lattice as long as it allows the establish of a long-range order at finite temperature.

We measure all the energies with respect to $W$ and all the times with respect to $W^{-1}$.

\section{Sudden quench dynamics}
\label{sec:quench}

\begin{figure}
  \includegraphics[width=\linewidth]{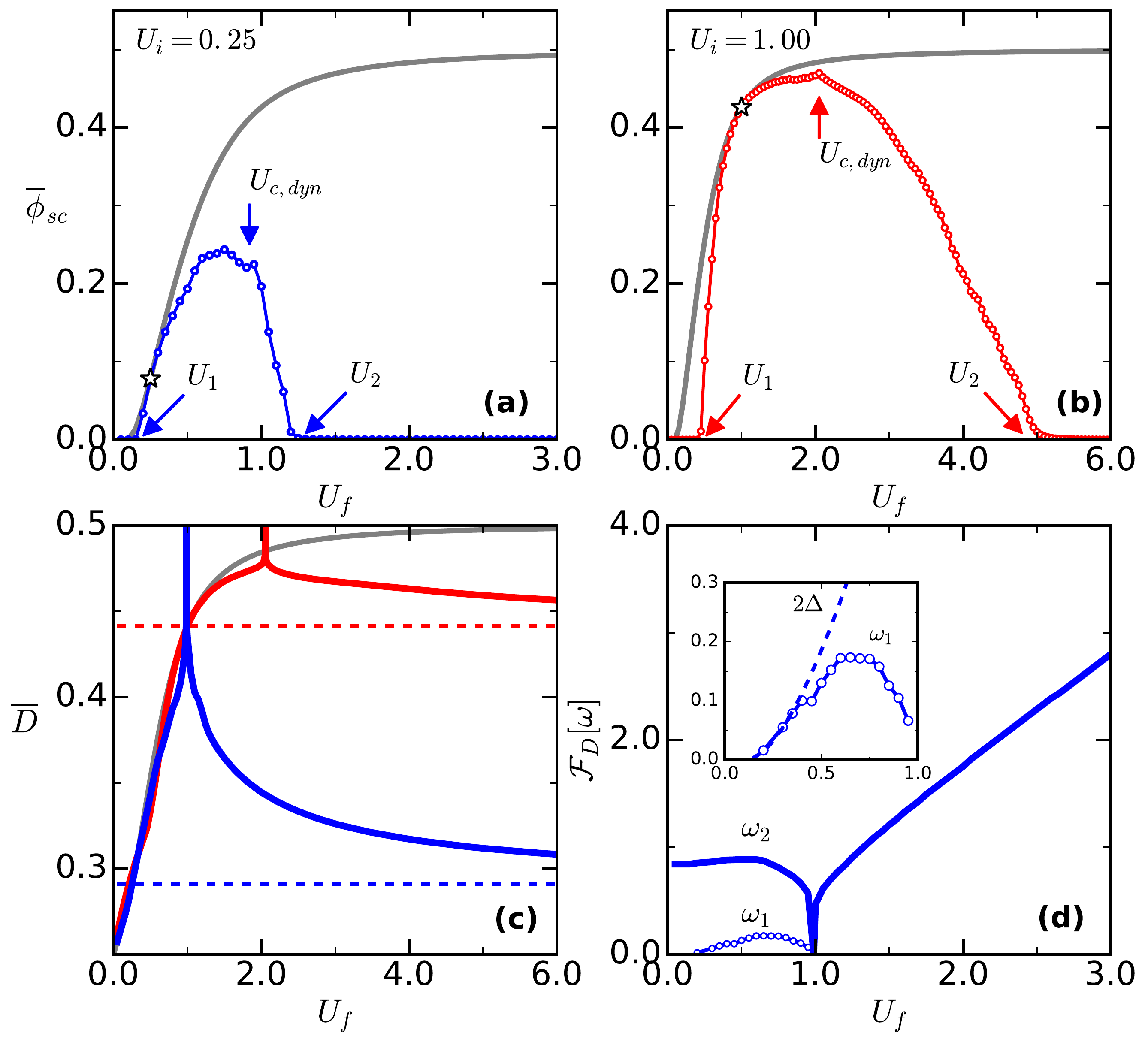}
  \caption{Sudden quench phase diagram. Top panels: (a)-(b) stationary values of the superconducting order
  parameter after a sudden quench as a function of the final value of the interaction $U_f$ and 
  for two initial values (indicated by stars) $U_i = 0.25$ (a) and $U_i = 1.0$ (b).
  The solid line represents the equilibrium value of the order parameter corresponding to $U_f$.
  The arrows indicate the positions of the symmetry restoration critical points $U_1$ and $U_2$ and 
  of the underlying weak-to-strong coupling dynamical transitions $\ucd$ (see text).
  Bottom panels: (c) stationary values of the density of doubly occupied sites as a function of the
  final value of the interaction $U_f$ for $U_i = 0.25$ (blue) and $U_i = 1.0$ (red).
  Dashed lines indicate the initial values $D_i$.
  A sharp non-analyticity point appears at the dynamical transition as
  $D \to 1/2$.
  Grey line: equilibrium values corresponding to $U_f$.
  (d) Main peaks of the Fourier spectrum of the double occupancies
  ${\cal F}_D$ as a function of $U_f$. 
  Only the case $U_i=0.25$ is shown. 
  Inset: blow up of the $\omega_1$ frequency compared to the
  superconducting gap at equilibrium (dashed  line). The two match in
  the limit $U_f=U_i$.
}
  \label{fig:quench_dynamics}
\end{figure}

At equilibrium the ground state of the Hamiltonian displays 
a finite superconducting order parameter for $U>0.$
At half-filling the superconducting order is degenerate with a charge
density wave instability. In the following we will constrain the
dynamics to superconducting states and neglect possible
competitions between the two orders.~\cite{sentef_cdw}

Fig.~\ref{fig:quench_dynamics} summarizes the main results of the
quench dynamics, corresponding to the limit $\tau \to 0$ in
Eq.~\ref{eq:ramp}. 
Due to the effective single particle description of the electron dynamics 
any generic time dependent observable ${\cal O}(t)$ 
usually displays undamped oscillations around stationary values~\cite{schiro2010}
which are extracted by means of long-time averages
\begin{equation}
  \overline{\cal O} = \lim_{t\to \infty} \frac{1}{t}
  \int_0^t d\tau \braket{\Psi(\tau)}{\cal O}{\Psi(\tau)}.
\end{equation}
Panels (a) and (b) report the stationary values of the order parameter
$\phi_{sc} = \langle\cc_{i\up} \cc_{i \down}\rangle$ as a function of 
$U_f$ and $U_i=0.25$ (a) and $U_i=1.0$ (b)
compared to the zero temperature equilibrium value.

%The time evolved state as an energy larger than the ground state of the
%final Hamiltonian and 
The stationary order parameter after the quench results always smaller 
than the corresponding ground state value of the final Hamiltonian as 
result of the fact that the quench corresponds to a finite injection of energy 
into the system.  
For final values of the interaction larger and smaller than the initial one, 
there exists two critical points depending on the value $U_i$, 
$U_{1} < U_i$ and $U_{2} > U_i$, for which the quench dynamics leads 
to the complete melting of the long range order.  

The melting of the superconducting order can be understood by the excess 
of energy inserted into the system by the quench. This  brings the
system in some high temperature state for which the symmetry is
restored. 
However, it has been shown  that both critical points display
non-thermal behaviors,~\cite{matteo_afm,tsuji_afm,werner_strong_afm} 
meaning that the order parameter remains finite even for values 
$U_f$ for which an effective temperature larger than the equilibrium
one is established.  

Different origins for these non-thermal critical points 
have been discussed for the weak ($U_1$) and strong ($U_2$) coupling cases.
At weak coupling a symmetry restoration due to dephasing has been
invoked,~\cite{matteo_afm,tsuji_afm}
while at strong coupling the stability of the symmetry broken state
has been related to the long life-time of the  
excitations induced by the quench 
at large values of $U$.~\cite{werner_strong_afm,balzer_prx}

The differences between quenches towards the weak and strong coupling limits 
are further highlighted by a sharp transition between two different dynamical
regimes occurring for $U_f=\ucd$ at intermediate values of the final
interaction $U_{1} < \ucd < U_2$.
This can be appreciated in the dynamics of the double occupancies
$D=\quave{n_{i\up}n_{i\down}}$ as a function of  the final
interaction.
Looking at the stationary values $\overline{D}$ (panel (c)), we identify 
a \emph{weak coupling quench} region $U_f< \ucd$ in which
$\overline{D}$ is renormalized according to the equilibrium
expectation at the final interaction ($\overline{D}$ increasing with
$U_f$) and a \emph{strong coupling quench} region in which the double
occupancies decrease with $U_f$ 
getting more and more  locked  to their initial value with eventually
$\overline{D}  \to D_i$ for $U_f \to \infty$.
For $U_f=\ucd$, the dynamics of the double
occupancies flows to the  value $D =1/2$ which is a fixed
point of the Gutzwiller dynamics.
A sharp transition is therefore identified by a non-analyticity 
point in the $\overline{D}$ vs $U_f$ phase diagram (panel (c)).

Physical insight on the different dynamical regimes comes from the analysis
of the frequency spectrum of the doublons dynamics ${\cal F}_D$ which gives the
characteristic energies of the doublons excitations created during the
quench.
Fig.~\ref{fig:quench_dynamics}(d) reports the main peaks of 
${\cal F}_D$ as a function of $U_f$.
In the weak quench side $D$ oscillates with two main frequencies $\omega_1$ and $\omega_2$
which can be related to characteristic energy scales of the system.
The larger frequency $\omega_2$ is of the order of the coherent quasi-particle 
bandwidth,
whereas $\omega_1$ can be related to the non-equilibrium superconducting gap.
This can be understood by looking at the inset of panel (d) where the frequency
$\omega_1$ is compared to the equilibrium gap $2\Delta$.
$\omega_1$ follows the dome-like behaviour of the  order parameter
and merges with the equilibrium value $2\Delta$  for $U_f\to U_i=0.25$.
In the strong quench region $U_f > \ucd$  only the mode $\omega_2$ 
survives and the frequency of the oscillations increases with $U_f$ as 
$\omega_2 \sim U_f$.

We can give a physical interpretation of the excitations induced by the 
quench in the various regimes.
In the weak coupling side these involve  the low-energy coherent
quasiparticle excitations, with the  emergence of a collective amplitude mode ($\omega_1$).
On the other hand only high-energy excitations,
approximatively located at the energies of the incoherent Hubbard
bands,  are activated  for $U_f > \ucd$.  
We can relate these excitations to local pair breaking processes that
frustrate the growth of the double occupancies despite the large value of the final 
attractive interaction.
The complete freezing of the excitations ($\omega \to 0$)
occurs at $U_f = \ucd$. 

For the repulsive AFM model the occurrence of such sharp crossover 
between two different dynamical regimes  has been identified with a dynamical 
transition~\cite{matteo_afm} similar to that observed 
in the paramagnetic case.~\cite{moekel_quench,eckstein_quench,schiro2010}
The presence of a dynamical transition can be related to the
corresponding equilibrium properties of the model.
In the paramagnetic case this has led to the
identification of the dynamical transition as the 
dynamical counterpart of the Brinkmann-Rice metal-to-insulator
transition.~\cite{schiro2010}
This has been further supported by the fact that the dynamical critical point
merges to the equilibrium Mott transition for an adiabatically slow
change of the interaction parameter.~\cite{sandri_ramps,eckstein_ramps} 

A similar analogy can be tried to be extended to the dynamics in the 
presence of long range ordered states.  In the AFM case the possibility 
of an \emph{incoherent antiferromagnet} has been proposed.~\cite{matteo_afm}
However, in the AFM case and in the equivalent
attractive superconducting case, this  cannot 
be related to any equilibrium counterpart. 
Indeed, in the symmetry broken phases, for both attractive and repulsive cases, 
the order parameter is finite for each value of
the interaction and no interaction driven transition is retrieved in
the zero temperature equilibrium phase diagram. 
We mention that, contrary to the paramagnetic case, 
the possible presence of a sharp transition
between two dynamical regimes in the presence of long-range order has
not been addressed  so far within DMFT. 
However, clearly distinct dynamical properties have been shown to hold 
for quenches at weak~\cite{tsuji_afm} and strong~\cite{werner_strong_afm} 
coupling, implying that a crossover between the two regimes is likely to 
take place at intermediate quenches.

It should be noticed that the equivalence between the repulsive and
attractive cases breaks down away from half-filling. In particular, in
the repulsive case the dynamical transition turns into a crossover at
any finite doping,~\cite{schiro2010} as related to the disappearance of the Mott
transition for $\langle n \rangle \ne 1$. On the contrary, the
metal-pairing insulator transition in the attractive case survives at
finite doping where, in infinite dimensions, it becomes of the first
order.~\cite{massimo_pairing_ins} Similarly, the dynamical
transition in the normal (non-superconducting) phase of the attractive
case survives  at finite doping. In this work we did not consider 
the study of the dynamical transition away from half-filling in the
superconducting phase. However  we expect that a similar  dynamical
transition may be retrieved for $\langle n \rangle \ne 1$.

\section{Finite ramp dynamics}
\label{sec:ramp}
The above dynamical properties show the emergence of stationary
regimes in which the dynamical properties are not fully
understood by a simple comparison of the quench and equilibrium phase
diagrams. 
The main goal of the rest of the paper is to investigate this connection by
looking at the evolution 
from the sudden quench to the adiabatic limit, where the phase
diagrams in  Figs.~\ref{fig:quench_dynamics}(a)-(b) are expected to 
merge with the zero temperature equilibrium one (grey lines).
In the following sections we will discuss such evolution by considering
the variation of the interaction parameter over a finite time scale.

\begin{figure}
  \includegraphics[width=\linewidth]{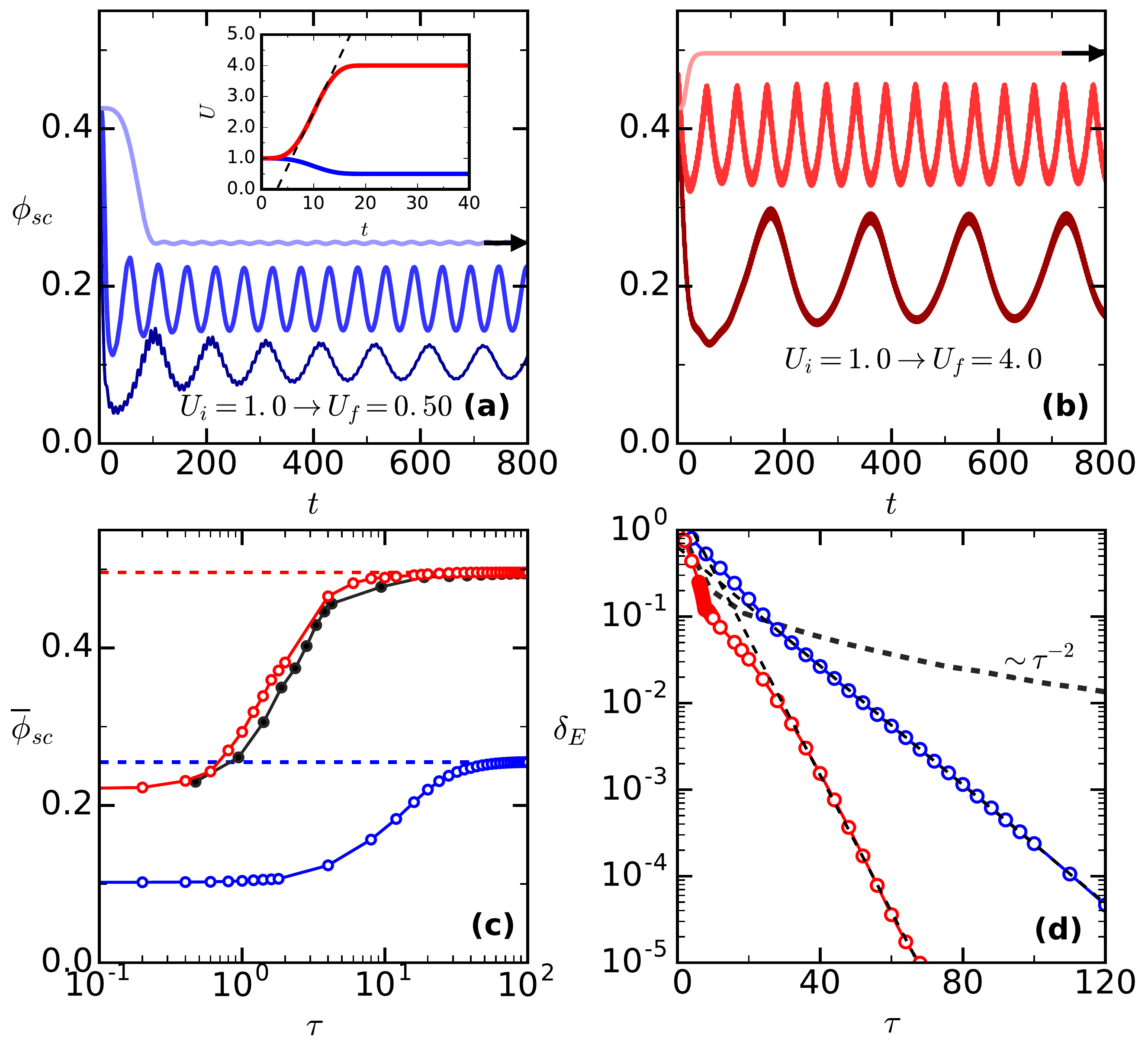}
  \caption{Adiabatic dynamics of superconducting states: 
    Top panels (a)-(b): dynamics of the superconducting order
    parameter for an interaction change from  $U_i=1.0$ to $U_f=0.5$
    (a) and $U_f=4.0$ (b) and different values of the ramping
    parameter. From bottom to top $\tau = 0,~12,~120$ and $\tau =
    0,~2,~30$  for (a) and (b) panel respectively.
    	The arrows indicate the ground state values for the final Hamiltonian.
    Inset: time dependent protocol for both values of $U_f$ and
    $\tau=20$. The dashed line indicates the linear approximation
    discussed in the text.    
    Bottom panels: (c) stationary value of the order parameter as a
    function of the ramping time $\tau$ for $U_f=0.5$ (blue) and
    $U_f=4.0$ (red). Black dots represent the results for $U_f=4.0$
    obtained with the linear approximation of the ramp. 
    (d) Adiabaticity parameter (see text), color code as in (c).
    Red dashed line represents the adiabaticity parameter proportional to $\tau^{-2}$ 
    for $U_f=4.0$ in the case of a linear ramp.   }
  \label{fig:ramp_dynamics}
\end{figure}

\subsection{Adiabatic dynamics of superconducting states}
\label{subsec:rampA}
We start considering the evolution of the non-equilibrium
superconducting states for finite variation rates of the 
attractive interaction. 
Figs.~\ref{fig:ramp_dynamics}(a)-(b) show the order parameter dynamics
for two interaction quenches, respectively in the weak and strong coupling side
of the quench phase diagram, $U_i=1.0$ to $U_f=0.5$ (a) and 
$U_f=4.0$ (b), and different values of the ramping parameter  $\tau$.  
Increasing the value of $\tau$ the  order parameters
become larger with respect to the sudden quench case. Eventually they
reach the equilibrium value, showing that the dynamics
approaches the  adiabatic limit for slow enough variations of the 
interaction.
The dynamics of the order parameter displays an undamped slow gap
amplitude mode  whose frequency increase as a function of $\tau$.
This reflects the  building up of a larger gap as the equilibrium ground
state is approached  in the adiabatic limit.

The time scale for the recovery of the adiabatic limit strongly depends
on the final value  of the interaction being shorter for the strong
coupling quench as compared to the weak coupling one. This can be
appreciated in the stationary values of the order parameters as a
function of $\tau$ (panel (c)).
For the quench towards the weak coupling limit ($U_f=0.5$) the
equilibrium value is reached for $\tau \gtrsim 50$ whereas in the
strong coupling case ($U_f=4.0$) this limit is almost reached  already
for  $\tau \gtrsim 15$.

The need for a shorter ramping time for the recovery of the equilibrium limit 
implies a smaller energy injection for fixed value of the ramping time. 
This fact can be understood on the basis of a Landau-Zener argument~\cite{Landau,Zener} 
in relation to the gap of the final Hamiltonian. In the attractive Hubbard model 
the gap is a monotonically increasing function of  the attractive interaction
and for a larger  the gap a smaller number of excitations is expected to
be created  during the continuous modification of  the Hamiltonian at
given variation rate. 

We measure the energy injected during the ramp $\tau$ with respect
to the sudden quench limit
\begin{equation}
  \delta_E(\tau) = \frac{E_f(\tau)-E_{eq}}{E_0-E_{eq}}.
\end{equation}
 $E_f(\tau) = E_{f} (t=\tau)$ is the energy after the ramp 
which stays constant for $t>\tau$. 
$E_0=E_f(\tau=0)$ is the energy in the sudden quench limit and
$E_{eq}$ is the zero temperature equilibrium energy corresponding to
the final value of the interaction.
For large ramping times $\delta_E$ decreases exponentially with $\tau$
on a scale that increases going from the weak to the strong
coupling regime. 

It is interesting to compare the above results with what obtained using a
linear ramp. As can be seen in the inset  of panel (a),  $r_{\tau}(t)$
can be approximated in the middle of the 
ramp $t = \tau/2$ with a line corresponding to $\tau'=\frac{3}{4\pi}
\tau$. Comparing the stationary values of the order parameter we see
that similar results are obtained for both protocols (full dots in
panel (c)), namely the
interpolation between the quench and the equilibrium values occurs for
ramping times of the same order.
On the other hand, the decay of the excitation energy changes for
larger values of 
$\tau$. In particular, for the linear ramp an expected power law
decay~\cite{polkovnikov_adiabatic} $\delta_E(\tau) \sim  \tau^{-2}$ is
retrieved (dashed line in panel (d)). This means that  at large
$\tau$ the linear approximation of $r_{\tau}(t)$ becomes less and less accurate and the
smooth variations of $r_{\tau}(t)$ for $t\simeq 0$ and $t \simeq \tau$ 
play a major role in determining the shorter ramping time needed to reach the 
adiabatic limit.
Despite such a difference,  the qualitative physics discussed in the rest of the  paper does not   
depend on the ramping protocol.   

%the fast dissipation of the energy injected by
%the quench.

\subsection{Adiabatic restoration of superconductivity}

We now investigate the adiabatic recovery of a finite order
parameter for final values of the interaction for which
the sudden quench dynamics leads to  melting of the 
superconducting order.  
As done before, we start from the initial interaction $U_i=1.0$ and
consider the ramp dynamics for $U_f=0.25$ and $U_f=6.0$, respectively
in the weak ($U_f<U_1$) and strong ($U_f>U_2$) coupling regions of the
sudden quench phase diagram (see Fig.~\ref{fig:quench_dynamics}(b)).

\begin{figure}
  \includegraphics[width=\linewidth]{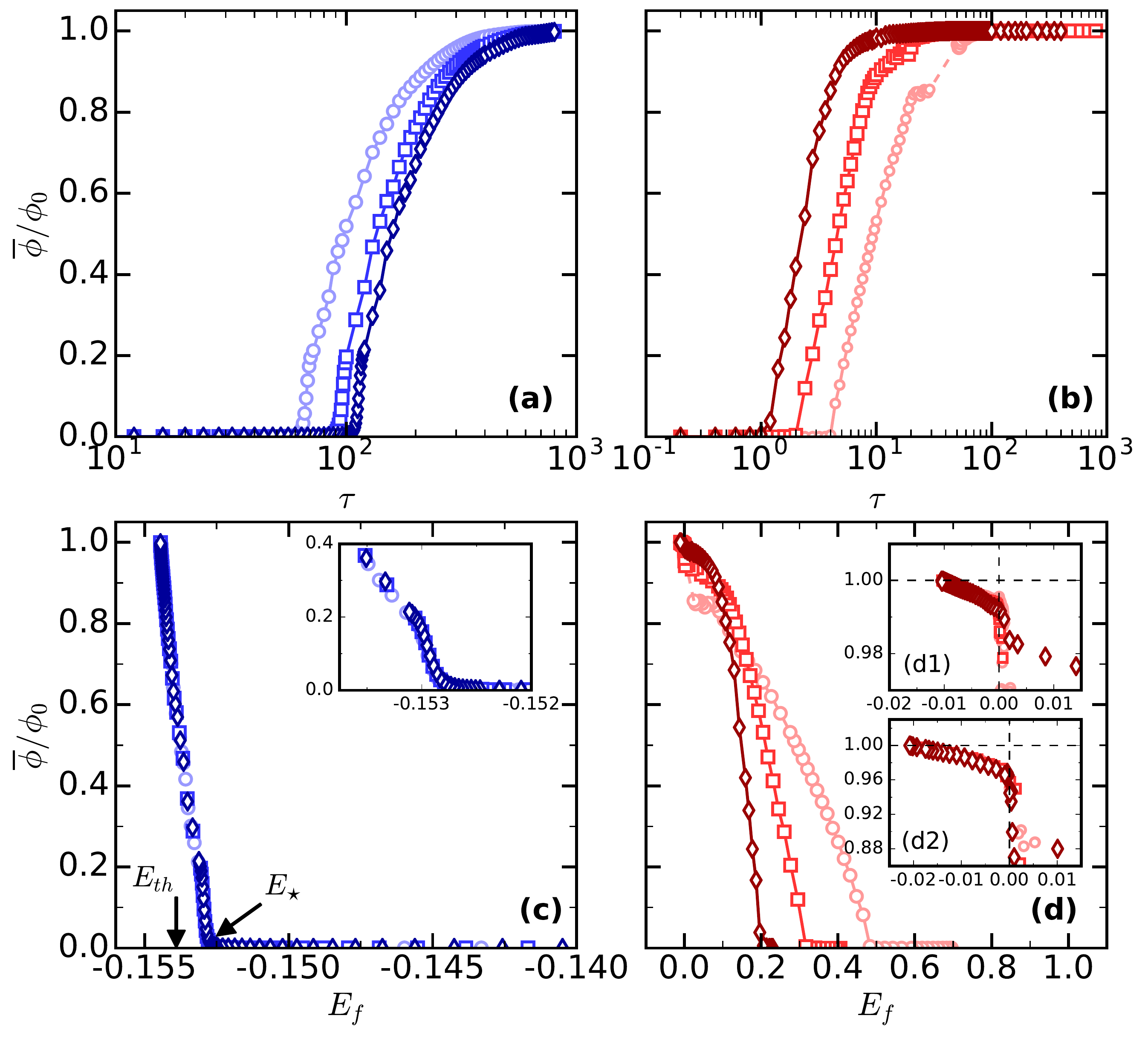}
  \caption{Adiabatic restoration of superconductivity. Top panels:
    stationary order parameters measured with respect to the
    equilibrium one at $U_f=0.25$ (a) and $U_f=6.0$ (b) as function
    of $\tau$ and different values of $U_i$. (a) $U_i=1.0$ (circles),
    $U_i=2.0$ (squares) and $U_i=3.0$ (diamonds). (b) $U_i=0.5$
    (circles), $U_i=0.75$ (squares) and $U_i=1.0$
    (diamonds). Bottom panels: stationary order parameters of
    panels (a)-(b) plotted as a function of the final energy after the
    ramp. Insets: (d1) Blow-up around $E_f=0$ of the data in panel (d). 
    (d2) Same of (d1) for a different set of $U_i$ and $U_f=3.0$. 
    $U_i=0.25$ (circles), $U_i=0.35$ (squares) and $U_i=0.45$ (diamonds).}
  \label{fig:symmetry_restoration}
\end{figure}

In Figs.~\ref{fig:symmetry_restoration}(a)-(b) we show that 
increasing the ramping time  the order parameter vanishes up to a
critical  value $\tau_{\star}$ above which superconductivity is
recovered. For $\tau > \tau_{\star}$ the order parameter increases and
eventually reaches the equilibrium value for $\tau \gg \tau_{\star}$.  
Similar features to that discussed in Fig.~\ref{fig:ramp_dynamics} are
retrieved, namely  the approach to the adiabatic  limit is
much slower in the weak coupling case compared to the strong coupling
one. 

The energy injected into the system and, in turn, 
the superconductivity restoration ramp  time $\tau_{\star}$ 
depends on the initial state and its overlap with the excited 
states of the final Hamiltonian. 
Considering quenches to the same final Hamiltonian and changing the
value of the initial interaction  $U_i$ we observe that
the superconductivity restoration ramp time $\tau_{\star}$ increases
as $U_i$ is, respectively, increased for $U_f=0.25$ and decreased for $U_f=6.0$.
This is expected by the fact that the larger is the difference
$\Delta U = \left| U_i  -  U_f \right|$ the larger is the energy injected into the system
and, for fixed final interaction, the slower is the variation rate
needed to recover the adiabatic limit.  

The relation between the injected energy during the ramp and the
dynamical restoration  of symmetry is further exploited in
panels (c)-(d), where the stationary values of the order parameter
obtained at each $\tau$ are plotted as a function of the final energy
after the ramp $E_f(\tau)$.  Two clearly distinct behaviors emerge for
quenches towards the weak  and the  strong coupling limits. 

In the weak coupling case ($U_f=0.25$) all the $\overline{\phi}(E_f)$ curves
collapse onto a single one showing that the time evolved state
approaches the steady state independently of the initial
conditions for given value of the energy injected during the ramp.
In particular, this reveals the existence of a universal value of the energy after the
ramp $E_f = E_{\star} \simeq -0.1528$ above which the symmetry is
restored.  

A completely different scenario is obtained in the strong coupling
case ($U_f=6.0$) where different $\overline{\phi}_{sc} (E_f)$ curves
are obtained for 
different values of the initial interaction. This shows that, contrary
to the previous case, the dynamics retains memory of the initial
state and, as a consequence, the energy 
threshold for the symmetry restoration is different for each value
$U_i$.

%, being larger the smaller $U_i$.
% We notice that, as anticipated in Sec.~\ref{sec:quench}, in both cases the 
% adiabatic recovery of a finite order parameter does not display a thermal
% behavior.  
% This straightforwardly derives from the initial state memory in the strong 
% coupling limit, while for the weak coupling case it can be seen by 
% comparing the energy threshold $E_{\star} $ with the internal energy
% corresponding to the equilibrium thermal transition, $E_{th} \equiv
% \text{Tr} (e^{-\beta_c {\cal H}_f} {\cal H}_f)/ \text{Tr} (e^{-\beta_c
%   {\cal H}_f})$ with $\beta_c = 1/T_{c}$ being $T_c$ the critical
% temperature of the Hamiltonian.
% As indicated by the arrow in Fig.~\ref{fig:symmetry_restoration} (c) $E_{th} < E_{\star}$.

We can try to extract from these results some information about the
properties of the stationary dynamics and the mechanisms for the
symmetry restoration. To do so we imagine to decompose the time
evolving state for $t>\tau$ in terms of eigenstates of the final
Hamiltonian 
\begin{equation}
  \ket{\Psi(t>\tau)} = \sum_n c_n e^{-i\epsilon_n t} \ket{\Psi^f_n},
\end{equation}
where $\mathcal{H}_f \ket{\Psi_n^f} = \epsilon_n \ket{\Psi_n^f}$
and the coefficient $c_n$, such that $\sum_n |c_n|^2=1$, are
determined by the dynamics during the ramp. 
Defining the expectation value of the the order parameter
operator onto a single eigenstate $\phi_{nn}$,~\footnote{The expectation value of
  the order parameter operator between two eigenstates is strictly
  zero \unexpanded{$\phi_{nm} =  \braket{\Psi_n}{ \cc_{i \up} \cc_{i
        \down}}{\Psi_m} = 0$}   for symmetry.
The proper definition in this case should take into account the long
range order \unexpanded{ $|\phi_{nm}|^2 = \lim_{|i-j| \to \infty}
\braket{\Psi_n}{ \cc_{i \up} \cc_{i \down} \ca_{j \down} \ca_{j \up} }{\Psi_m} $}}
the stationary value of the order parameter is described by
the so called diagonal ensemble\footnote{We assume here that there are no degeneracy in the spectrum.}
\begin{equation}
  \overline{\phi}=\sum_{n} |c_{n}|^2 \phi_{nn} .
\end{equation} 
In the weak coupling case the collapse onto a single curve of all the $\overline{\phi}(E_f)$
makes it reasonable to assume that the decomposition of the time evolved
state is narrowly peaked in energy around the value of the final
energy after the quench $E_f$ and that the expectation values 
smoothly varies in energy, as predicted by the Eigenstate Thermalization
Hypothesis (ETH)~\cite{srednicki_ETH,Rigol_ETH}
and its extension to symmetry broken states.~\cite{srednicki_ETH_SBS}
 Under this assumption the existence of a universal energy $E_{\star}$ for the symmetry restoration
shows, as originally proposed in Ref.~\onlinecite{giacomo_michele_2012}, the
existence of an energy edge in the spectrum  of the final Hamiltonian
which separates a subspace containing  symmetry breaking eigenstates
for  $E<E_{\star}$ from a subspace  containing \emph{only} symmetry
y
invariant ones  for $E>E_{\star}$. 
The dynamical restoration of symmetry for $E>E_{\star}$ is then
due to the fact that the time evolving state overlaps only with
symmetry invariant eigenstates  so that the order parameter dephases
to zero in the long time limit.

The dependence only on the energy $E_f$ may suggest a thermal origin
for the  restoration of symmetry. However, by comparing
the energy threshold $E_{\star}$ with the internal energy 
corresponding to the equilibrium thermal transition, $E_{th} \equiv
\text{Tr} (e^{-\beta_c {\cal H}_f} {\cal H}_f)/ \text{Tr} (e^{-\beta_c
  {\cal H}_f})$ with $\beta_c = 1/T_{c}$ the inverse of the critical
temperature of the final Hamiltonian, we find that this is not the
case being $E_{th} < E_{\star}$ (see arrows in
Fig.~\ref{fig:symmetry_restoration} (c)). 
In general, this reflects the fact that the Gutzwiller dynamics is not
expected to describe the evolution towards a true thermal state.
%  the pre-thermal character of the 
% dynamics described by the Gutzwiller  ansatz which, as already
% mentioned, is not expected to describe the  evolution towards  a 
% true  thermal state.  
At the same time, we notice that the existence of an energy edge in
the spectrum does  
not necessarily imply its equivalence with $E_{th}$. Indeed, it may
happen that the thermal transition occurs at $E_{th}<E_{\star}$ due to the
larger entropic contribution of the symmetry invariant
states.~\cite{giacomo_michele_2012}   

The same analysis does not apply in the strong coupling regime where
the memory of the initial state  clearly highlights the breakdown of
any thermal behavior.
% Contrary to the previous case, this cannot be related to the failure 
% of the variational dynamics to describe thermalization.
Two different scenarios can be considered.
One possibility is that the ETH breaks down for the final Hamiltonian.
In particular, the $c_n$ decomposition is peaked around the
energy $E_f$ and the expectation values have large fluctuations for
eigenstates close in energy,
so that for different choices of $c_n$ different 
expectation values in the diagonal ensemble are obtained.
% this reflects the breakdown of the ETH
% hypothesis.
%This appears as an intrinsic property of the time evolving state and, 
% The variational dynamics does not allow to better
% investigate the causes.
% a breakdown of the thermalization is highlighted by the dynamics
% dependent on the memory of the initial state. 
% In particular, it may occur that, while a finite overlap with symmetry
% breaking states is  retained up to $E_{\star}$ during the dynamical
% evolution of the initial state, the thermal transition may occur already
% for smaller internal energies due to the entropic contribution of the
% symmetry invariant states.
% We may apply similar arguments for the description of the
% symmetry restoration in the strong coupling case.
On the other hand, if we assume ETH to hold we conclude that  the
decomposition of the  time evolved wave function  has to be widely
spread in energy.   
In that case, assuming as before the existence of an energy edge
$E_{\star}$  in the spectrum, for a given initial state a finite
overlap with the  symmetry  broken states may be retained also for
energies $E_f>E_{\star}$.   
Therefore the symmetry restoration can be understood as due to 
a spectral transfer from the low to the high energy part of the spectrum
which strongly depends on the configuration of the initial state.
While both are possible scenarios, a thorough analysis of the
spectrum is clearly beyond the description of the present variational
approach. 

We observe that the strong dependence of the dynamics on the initial
state might in principle suggest the absence of any adiabatic
behavior. Indeed, if the stationary dynamics depends on both the
energy pumped into the system and the initial state one could expect
the convergence to different states in the long ramping time limit.  
Nonetheless, as seen in panel (b) the dynamics at strong coupling are all
converging to the ground state of the final Hamiltonian, showing that the
adiabatic limit is eventually recovered.
In the following we investigate how the recovery of an adiabatic
dynamics in the strong coupling case takes place for larger ramping
times  or, equivalently, smaller injected energies $E_f$.

\begin{figure}[t]
  \includegraphics[width=\linewidth]{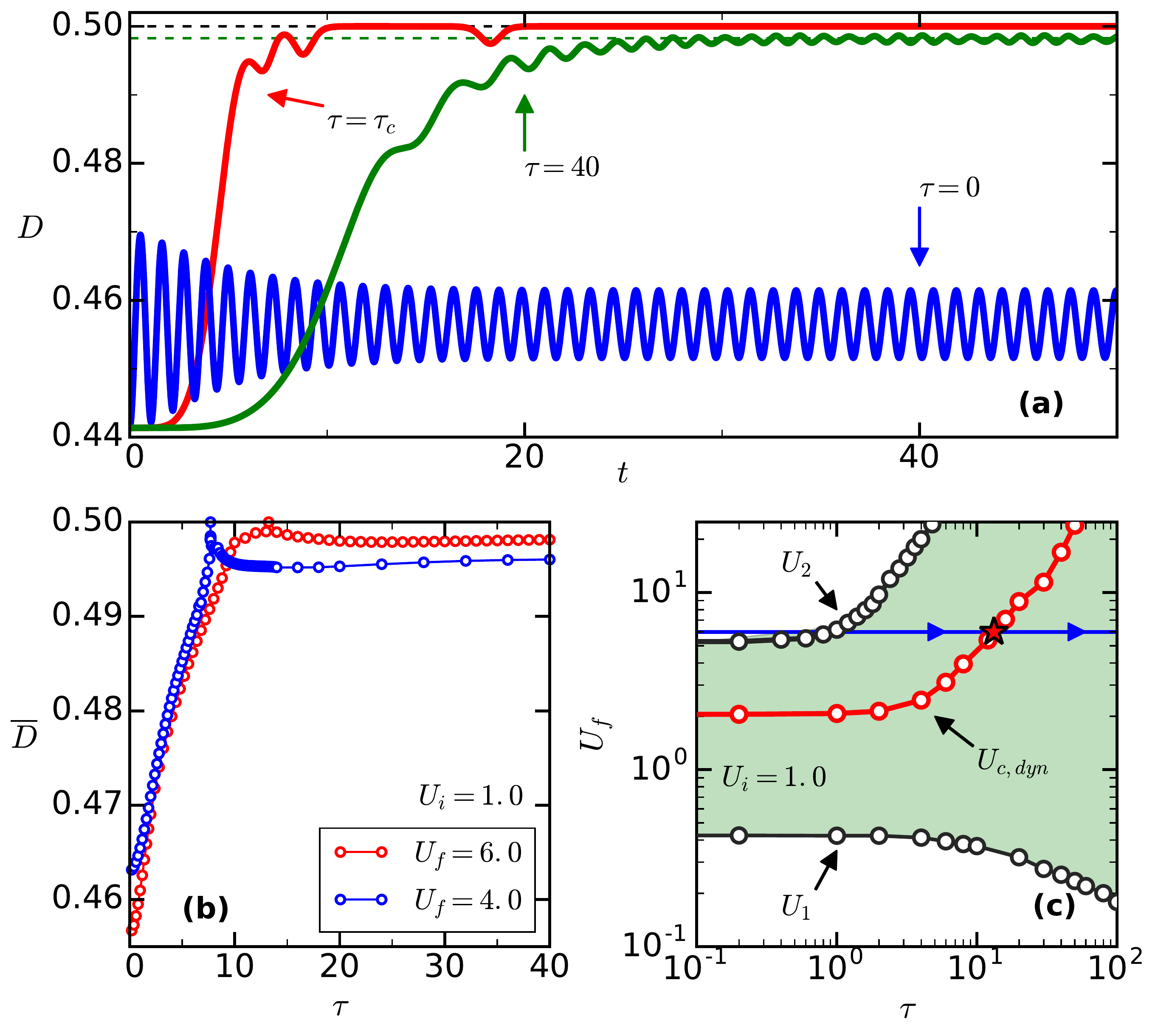}
  \caption{Dynamical transition as a function of $\tau$. 
    (a): Dynamics of the double occupancies for $U_i=1.0 \to U_f=6.0$ and 
    $\tau=0$ (blue line), 
    $\tau=40$ (green line) and $\tau=13.2294$ (red line).
    Black and green dashed lines indicate respectively the 
    values $D=1/2$ and the equilibrium value corresponding to $U_f$.
    (b): Stationary values of the double occupancies as a function
    of $\tau$ for $U_i=1.0 \to U_f=6.0$ (red) and $U_i=1.0 \to U_f=4.0$ (blue).
    (c): Dynamical phase diagram as a function of $\tau$. 
    The symmetry restoration points $U_1$ and $U_2$ mark the
    boundaries in which a finite order parameter is observed (shaded
    area). Red points indicate the values of $\ucd$ as a function of
    $\tau$. The horizontal arrows and the star indicate the crossing of the
    dynamical transition from the strong to the weak coupling region
    as a function of $\tau$ discussed in panels (a) and (b).}
  \label{fig:ramp_dynamical_transition}
\end{figure}

\subsection{Recovery of the adiabatic limit and strong-to-weak coupling dynamical transition}
Inset (d1) in Fig.~\ref{fig:symmetry_restoration} shows the blow-up around the ground state energy of the
final Hamiltonian $ E_f \approx E_{eq}$ for the $\overline{\phi}(E_f)$
curves in panel (d).
We observe that the adiabatic behavior is abruptly 
recovered as soon as the the energy crosses the value $E_f=0$, 
after which the collapse of all the different $\overline{\phi}_{sc}(E_f)$ 
curves onto a single one is retrieved.
Choosing a different value of $U_f$ in the strong quench regime and 
a different set of $U_i$ (inset (d2) in Fig.~\ref{fig:symmetry_restoration}) 
we can see that the recovery of the collapse of the $\overline{\phi}_{sc}(E_f)$ for $E_f <
0$ is a universal feature of quenches in the strong coupling regime.
This fact highlights a clear link between the two different dynamical behaviors
and the weak-to-strong quench dynamical transition discussed in
Sec.~\ref{sec:quench}.

Within the Gutzwiller ansatz the energy $E=0$ is realized only for $D=1/2$ which, 
as already mentioned in Sec.~\ref{sec:quench},
is the fixed point of the dynamics defining the transition between 
the weak and the strong coupling quench dynamics.
In Fig.~\ref{fig:ramp_dynamical_transition}  we show that 
such transition can be crossed as a function of the ramping time
from the strong to the weak coupling side.
In particular, in panel (a) we see that for the
quench $U_i=1.0 \to U_f=6.0$ (corresponding to the diamonds in
Fig.~\ref{fig:symmetry_restoration}(d)) the fixed point of the
dynamics  is reached for the critical value $\tau_c \simeq
13.2294$, for which, after  the value $D=1/2$ is reached,  the
dynamics stays constant apart  
from a small bump at $t\approx 18$ which is related to the numerical uncertainties 
in exactly reaching the fixed point.

In the quench limit ($\tau=0$) the signatures of the strong coupling
dynamics are visible, namely the double occupancies are locked around
a value close to the initial one with fast oscillations of frequency
$\sim U_f$. An almost adiabatic dynamics is established for $\tau=40$ where 
the double occupancies reach the equilibrium value with small residual oscillations. 
A non-analyticity point in the stationary value of the double occupancies is 
obtained for $\tau=\tau_c$ (panel (b)) which, similarly to what discussed 
in Sec.~\ref{sec:quench}, defines the  crossing as a function 
of the ramping time $\tau$ of the dynamical transition from the strong 
($\tau<\tau_c$) and the weak ($\tau>\tau_c$) coupling quench dynamical 
regimes. This corresponds respectively to  $E_f(\tau<\tau_c)>0$ and $E_f(\tau>\tau_c) < 0$,
thus showing that the transition between the weak and the
strong quench dynamics separates an adiabatic regime in which the
dynamics depends only on the energy injected into the systems 
and a regime for which a dynamics 
strongly dependent on the initial state sets in.

In Fig.~\ref{fig:ramp_dynamical_transition}(c) we report the evolution
of the dynamical phase diagram as a function of the ramping time.
At fixed value of $\tau$ we define the two critical
interactions $U_1(\tau)$ and $U_2(\tau)$ at which the symmetry is
restored and the crossing of the dynamical transition $\ucd(\tau)$  as
the value of the final  interaction at which the 
non-analyticity in the double occupancies occurs.
In the adiabatic limit ($\tau \to \infty$), where the quench phase diagram converges to
the zero-temperature equilibrium one (see
Fig.~\ref{fig:quench_dynamics}), we expect $U_1(\tau) \to 0$ and
$U_2(\tau) \to \infty$. This is correctly reproduced in
Fig.~\ref{fig:ramp_dynamical_transition}(c) where $U_2$ is an
increasing function of $\tau$ and $U_1$ is slowly vanishing at large
$\tau$.  

At intermediate values $\ucd(\tau)$ clearly follows the increase of
$U_2$. Due to the absence of any interaction driven transition in the 
zero temperature equilibrium phase diagram we do not expect
$\ucd(\tau)$ to saturate at a finite value for $\tau \to \infty$. We thus conclude that
also $\ucd(\tau) \to \infty$ for $\tau \to \infty$, namely
the strong coupling region disappears in the adiabatic limit and  only
the fixed point survives   corresponding to the equilibrium atomic
limit.  
As a consequence at fixed value of $U_f > \ucd(\tau=0)$ the
recovery of the adiabatic limit always implies the crossing of the
strong-to-weak coupling quench dynamical transition (blue arrow in
panel (c)). 
This shows that the non-equilibrium states at strong coupling lack of
an adiabatic connection to the equilibrium ground state, meaning that
they cannot be adiabatically  transformed into the equilibrium ground
state at finite $U$ without entering the weak coupling side of the
transition.

\begin{figure}
  \includegraphics[width=\linewidth]{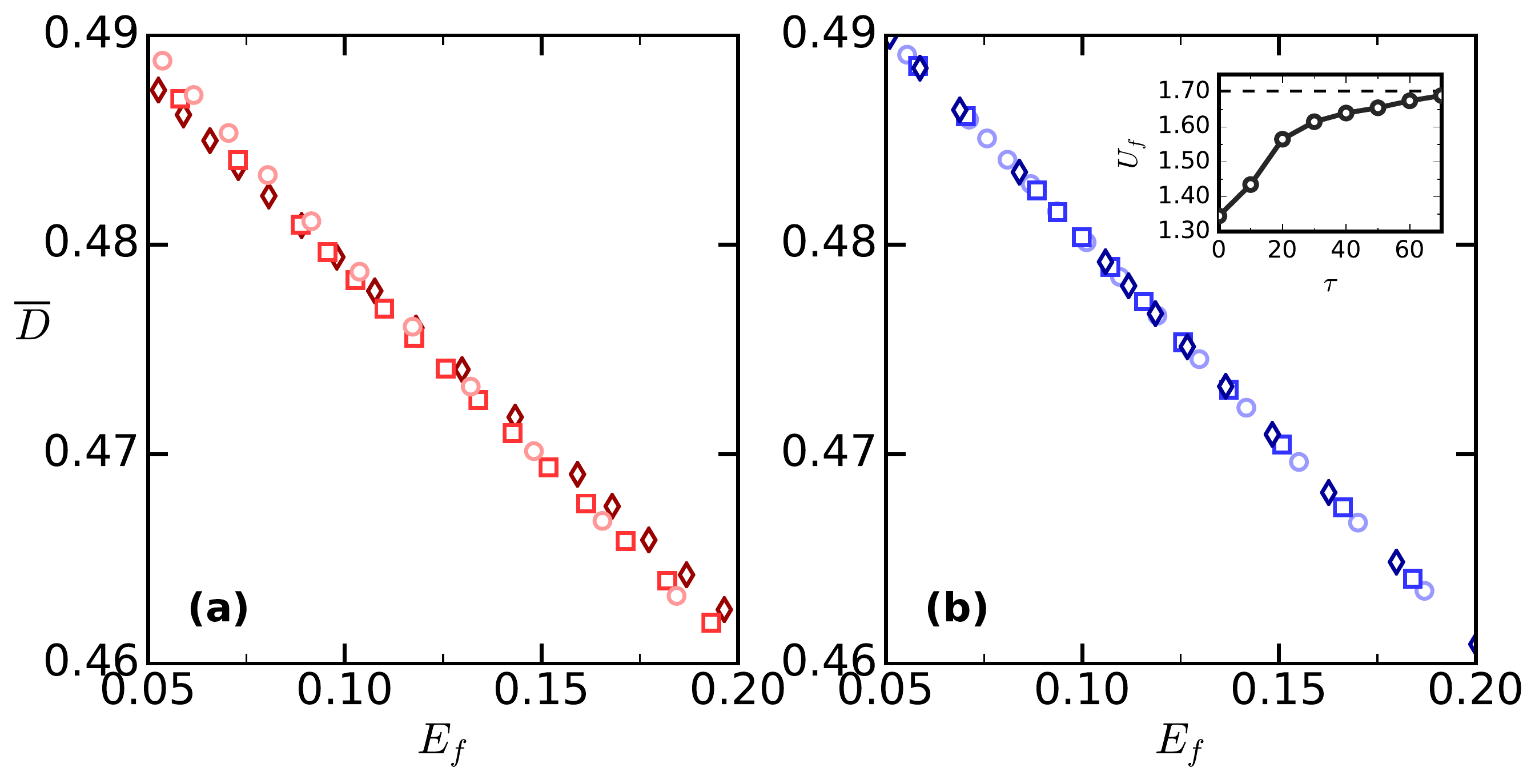}
  \caption{Stationary values of the double occupancies 
    for the superconducting (a) and normal phases (b) 
    as a function of the energy injected during the 
    ramp for $U_f=6.0$ and different values of $U_i$. 
    $U_i=1.0$ (circles), $U_i=0.75$ (squares) and $U_i=0.5$
    (diamonds). Inset: $\ucd$ as a function of $\tau$ for the normal
    phase. The dashed line indicates the equilibrium Mott transition.}
  \label{fig:sc_vs_normal}
\end{figure}

\subsection{Comparison to the normal state dynamics}
The last result suggests a relation between the lack of an adiabatic
connection to the equilibrium limit for the strong coupling case and
the initial state dependent dynamics in
Fig. ~\ref{fig:symmetry_restoration}(d). 
We check this by comparing the dynamics in the strong coupling regimes 
in both superconducting and normal (non-superconducting) phases
for which an intrinsically different scenario is expected.
In the normal phase the weak-to-strong quench dynamical transition 
converges to the metal-pairing insulator Mott transition $U_{\text{Mott}}$ 
for $\tau \to \infty$ (see inset in Fig.~\ref{fig:sc_vs_normal}(b))
and, contrary to the previous case, the non-equilibrium states for 
$U_f>U_{\text{Mott}}$ can be adiabatically transformed into the equilibrium 
pairing insulator without crossing any dynamical transition. 

To compare the dynamics in the two phases we look at the dynamics of
the double occupancies varying the ramping times and starting from different initial 
states. Fig.~\ref{fig:sc_vs_normal} reports the stationary values 
as a function of the energy after the ramp.

In the superconducting phase the same dependence on the initial state
shown in Fig.~\ref{fig:symmetry_restoration} is reproduced. On the
contrary the clear collapse onto a single curve is observed in the 
normal phase, for which a dynamics determined only by the final
Hamiltonian is obtained also in the strong quench regime.
This supports the fact that  the initial state dependence in the
superconducting phase is a manifestation of the  lack of an adiabatic
connection to the final ground state in the strong coupling quench
regime.

\section{Conclusions}

We used the time-dependent Gutzwiller approximation
to investigate the dynamics in the attractive Hubbard model.
In particular, we focused on the connection between the sudden quench limit
and the adiabatic dynamics following a slow variation of the attractive interaction.

The quench phase diagram displays two dynamical critical
points at both weak and strong coupling at which the initial
superconducting order parameter is melted after the quench
and a sharp transition between a weak and a strong coupling 
quench dynamical regimes related to the interplay between the delocalized (quasiparticles)
and the localized (Hubbard bands) degrees of freedom excited 
by the quench.

By changing the rate of variation of the attractive interaction
we showed that the adiabatic limit is reached for sufficiently slow 
ramps. The ramping time for reaching the adiabatic limit 
depends on the final Hamiltonian
and  it decreases going  from the  weak to the strong coupling limit.

In the case of final  values of the interaction for which
the sudden quench dynamics leads to the complete melting of 
the superconducting order, we demonstrated the existence of a minimum 
ramping time for the recovery of a finite order parameter. 
For fixed final Hamiltonian the minimum ramping time depends 
on the initial state of the system and two distinct behaviors 
appear for quenches in the weak and strong coupling limit.

At weak coupling a dynamics depending only on the 
energy injected during the quench is realized independently of the initial state. 
On the contrary strong coupling quenches are characterized by a 
dynamics which retains memory of the initial state.

We showed that this is closely related to the weak-to-strong coupling 
quench dynamical transition which, for fixed final Hamiltonian, can be crossed 
as a function of the ramping time. 
In particular, the dynamics dependent on the initial state in the 
strong coupling side suddenly collapses onto the initial state independent 
dynamics as the transition to the weak coupling side is crossed.
 
By tracking the evolution of the dynamical phase diagram
as a function of the ramping time and by a comparison to the dynamics in the
normal phase we argued that the initial state dependent dynamics is the result of 
the lack of an adiabatic dynamics for the strong coupling quench regime 
in the superconducting phase.

\section{Acknowledgments.}
I thank  M. Schir\'o, M. Fabrizio and A. Georges  for suggestions and
insightful discussions.  The research leading to these results has received
funding  from the European Research Council under the European
Union's Seventh Framework Programme (FP7/2007-2013) / ERC Grant
Agreement nr. 319286 (Q-MAC).

\bibliography{biblio_neq}

\end{document}